\documentstyle[insert]{article}
\input epsf

\newcommand{\beq}{\begin{equation}}
\newcommand{\eeq}{\end{equation}}

\large

\begin{document}

\title{\bf Implications for the Cosmic Ray Spectrum of a Negative Electron
Neutrino (Mass)$^2$}

\author{Robert Ehrlich\\George Mason
University\\Fairfax, VA 22030\\rehrlich@gmu.edu}
\large

\maketitle
\nopagebreak
\begin{abstract}
The hypothesis that the electron neutrino is a tachyon with $|m| 
\equiv\sqrt{-m^2}\approx 0.5 eV/c^2$ is consistent with certain
properties of the observed cosmic ray spectrum, including: the existence 
of a change in power law (the ``knee") at $E\approx 4\times 10^{15} eV$, 
the $E^{-3}$ power law after the knee, another change in power law (the 
``ankle") at $E\approx 10^{19} eV$, the changes in composition above the 
knee, the change in anisotropy at the knee, and the absence of a GZK 
cutoff above$E\approx 4\times 10^{19} eV$.  The hypothesis predicts a 
substantial flux of cosmic ray neutrons in a narrow energy region just 
above the knee of the spectrum.\\ \\

\end{abstract}

\newpage

\large
\section*{Introduction}


Stimulated by the recent report of neutrino oscillations,\cite{Oscill} 
which demonstrates that at least one type of neutrino has a non-zero 
mass squared, and persistently negative values for ${m_{\nu_e}}^2$ from 
tritium beta decay experiments,\cite{PDG} we conjecture that the 
electron neutrino is a tachyon with $m_{\nu}^2 \approx -0.25 eV^2/c^4,$ 
or $|m_{\nu}|\equiv\sqrt{-m^2} \approx 0.5 eV/c^2,$ and we consider the
predicted consequences for the high energy cosmic ray spectrum:

\begin{enumerate}

\item The cosmic ray spectrum should have a discontinuity in power law 
(a ``knee") at $E \approx 4\times 10^{15} eV$.

\item The power law for the spectrum should steepen by 0.33, i.e., go 
from $E^{-2.67}$ to $E^{-3.0}$ after the knee, and up to $E \approx 10^{18} eV$

\item The spectrum ($\times E^3$) should have a dip at $E\approx
10^{19} eV$, after which it should rise.

\item The average atomic mass of cosmic rays as a function of their 
energy should become increasingly heavy after the knee, but only up to 
$E \approx 10^{18} eV$, after which it should become increasingly light
over the ensuing decade, becoming almost pure protons at $E \approx 
10^{19} eV$

\item The variation of anisotropy of the cosmic rays with energy should 
show a change in slope at $10^{15} eV$.

\item Cosmic rays with energies in excess of $E\approx 4\times 10^{19} 
eV$ should be observed, despite the existence of a conjectured GZK 
cut-off around this energy\cite{Greisen} 

\end{enumerate}


The model can be thought of as an {\it example} of how the hypothesis of 
``tachyonic" electron neutrino might provide an explanation of the 
preceding regularities, some of which have more conventional 
explanations.  The hypothesis is consistent with other neutrino 
observations, and it makes a specific prediction: the existence of a 
cosmic ray neutron flux in a narrow range of energies just above the knee of
the spectrum. Therefore, it should not be dismissed out of hand, even if 
it conflicts with much conventional wisdom about the cosmic rays, and 
the nonexistence of tachyons.

\section*{Tachyons}

Tachyons, first postulated in 1967, by Bilaniuk, Deshpande, and
Sudarshan,\cite{Bilaniuk} and later by Feinberg\cite{Feinberg} have not
been taken seriously by most physicists, because of the paradoxes they
create, and because all experiments specifically\cite{tritium} searching 
for tachyons have turned up negative.\cite{Alvager,Baltay}  Thus, even 
though some of the equations of theoretical physics -- especially in
the field of string theory -- have implied the existence of tachyons, 
they have generally been eliminated from most respectable
theories.\cite{Gliozzi}

Whatever one's view of tachyons, their existence is clearly an 
experimental question. Those negative experiments to date merely rule 
out: (1) charged tachyons, (2) tachyons whose mass$^2$ is too close to 
zero to be resolved in the experiment, or (3) tachyons not produced or 
detected in sufficient abundance to be observed.  Weakly interacting 
neutral tachyons of low mass would have probably escaped detection, or 
else not be recognized as tachyons.  In
fact, Chodos, Hauser and Kostelecky suggested in 1985 that one or more 
of the three flavors of neutrinos is a tachyon.\cite{Chodos85} They 
based this idea on the fact that the best value of the square of the 
electron neutrino mass from the endpoint of the beta decay spectrum of 
tritium was at that time found to be negative by over two standard 
deviations.

In several papers Chodos et al.\cite{Chodos92,Chodos94} suggested that
one could test whether neutrinos are tachyons using one of their strange 
properties, i.e., that particle decays producing tachyons which are 
energetically forbidden in one reference frame are allowed in another.  
Thus, consider the energetically forbidden ``decay": $p\rightarrow
n+e^{+}+\nu_e$. For the decay to conserve energy in the rest frame of
the proton, the energy of the neutrino obviously would need to be 
negative in that frame.  (The
positron or neutron could not have negative energy in this frame, 
because then they would have negative energy in all frames.) However, 
tachyons, unlike other particles, have $E<p$ so they can change the sign 
of their energy when boosted to a sufficient velocity.  Thus, the 
tachyon energy in the proton rest frame, $E$ has the opposite sign from 
its energy in the lab $E_{lab} = \gamma(E - \beta p\cos\theta)$ when $\beta$
exceeds $E/p\cos\theta < 1$.  At the threshold energy for proton beta decay
the configuration of the three final state particles is a tachyon going
directly forward in the lab with zero energy (but zonzero momentum 
$p=|m|$), and a neutron and positron going directly forward with a 
common velocity.

The threshold energy for protons to decay is
found by making $E_{\nu}$ the least negative it can be in the CM frame, 
i.e., $-E_{\nu} = m_n + m_e - m_p \equiv \Delta$.  At threshold we take 
$\cos\theta = 1,$ so for a relativistic neutrino, 
$\beta_{th}=E_{\nu}/p_{\nu} \approx 1 + \frac{1}{2} m_{\nu}^2/{E^2}_{\nu}$, and
hence $\gamma_{th}=(1-\beta_{th}^2)^{-1/2}=\Delta/|m_{\nu}|$, so that

\beq
 E_{th} = \gamma_{th}m_p = \frac{m_p\Delta}{|{m_{\nu}}_e|}=
\frac{1.7\times10^{15}}{{|m_{\nu}}_e|}eV
\eeq

(For nuclei of mass number A, $m_p$ in the preceding formula is the mass 
of the parent nucleus, and $\Delta$ is the mass difference: $m(A,Z\pm 1) 
+ m_e - m(A,Z)$. As various authors have explained, the idea of 
``stable" particles decaying is less paradoxical if one reinterprets the 
emitted (positive energy) neutrino in the lab frame to be an absorbed 
(negative energy) neutrino from a background sea in the proton rest 
frame -- the so-called ``reinterpretation 
principle."\cite{Bilaniuk,Feinberg,Chodos85} In any case, the net result 
is that protons or other ``stable" nuclei at sufficiently high energies 
will undergo beta decay if the neutrino is a tachyon.

In order to test the prediction of Chodos et al. as applied to the 
cosmic ray spectrum, we need to calculate the phase space for proton or 
other stable nuclei to beta decay as a function of their energy.  This 
is done by integrating that small region of phase space in the CM (the 
parent proton or stable nucleus rest frame) for which the neutrino 
energy changes sign between the CM and lab frames.  The small fraction 
of the entire $4\pi$ solid angle that is available for any given 
(negative) neutrino energy in the CM is given by

\beq
\frac{\Delta\Omega}{4\pi} =
\frac{1}{2}\left(\frac{m_p}{E_{th}}\right)^2\left(1-x^{-2}\right),
\eeq

where $x=E/E_{th}.$  The results of
the phase space integration (done approximately using the relativistic 
approximation for neutrinos and electrons) give us the predicted mean 
free path before a proton undergoes beta decay.  For large $x$
the result is

\beq
mfp = \frac{90 eV^3 ly}{|m_{\nu}|^3x^2}
\eeq

\section*{Predicting the Cosmic Ray Spectrum Above the Knee}

We now describe how the assumption of a tachyonic neutrino can be used to
make specific predictions about the cosmic ray spectrum beyond $10^{15} 
eV$.  The input to the calculation are assumptions for: (1) the electron neutrino mass,
(2) the energy spectrum and composition of cosmic rays at their
source, and (3) the spatial distribution of sources.  For the
spatial distribution, we take an admixture of two source populations:
``near" and ``far" sources.  Near sources are assumed to create cosmic
rays that have path distances to Earth in the range $10^4$ to
$2\times 10^6$ ly, and far sources are assumed to have path distances to 
Earth in the range $2\times 10^6$ to $10^8$ ly.  (Clearly the terms 
``near" and ``far" are something of a misnomer here: in view of the
nonlinear paths of cosmic rays, their path lengths
can greatly exceed the distances to sources except at the highest 
energies.)

For the source spectrum we use an $E^{-2.67}$ power law that fits the 
spectrum up to $10^{15} eV$ -- see figure 2.  Essentially, we assume
that the source spectrum is $E^{-2.67}$ at all energies, and that any 
changes in the observed spectrum are due to particles in a given energy 
bin being shifted to lower energies as a result of beta decay.  We have 
ignored the possibility that the spectra of sources might be distance 
dependent due to evolution.  Since the composition of the cosmic ray 
spectrum above the knee is not known very well, we try various 
compositions and see which is in best agreement with the data.  Again, 
however, we assume a fixed composition ratio independent of energy at 
the source, and assume that variations in composition with energy arise 
as the successive thresholds are reached for nuclei in the spectrum to 
decay.  (Heavy nuclei tend to have higher thresholds than lighter ones 
because of the proportionality of $E_{th}$ on the nuclear mass.)

The case of the helium nucleus component in the cosmic rays is a special 
situation, because no beta decay has an alpha particle in the
final state.  The $E_{th}$ in the case of $He^4$ cosmic ray particles is
the threshold energy at which the energetically forbidden decay: $He^4 
\rightarrow 2p + 2n +\nu$ becomes possible for tachyonic neutrinos.  
(Equation 1 still applies here, with $\Delta = 2(m_p + m_n) - m_{He^4},$
and $m_p$ is the mass of the parent alpha particle.)

To derive the curves shown in the figures the Monte Carlo method was 
used.  Protons and nuclei were generated at various distances from Earth 
according to the assumed $E^{-2.67}$ source spectrum, and the fate of all
particles in a given energy bin was considered to be the same, as their 
progress toward Earth was followed. For protons leaving sources
above the threshold energy for decay, there is a chain of decays 
$p\rightarrow n\rightarrow p\rightarrow n\rightarrow p\cdots$ which
stops when the nucleon either reaches Earth or else has its energy 
reduced below threshold.  (We assume that at the source it was a matter
of random choice as to whether nucleons
start out as neutrons or protons.)  A similar decay chain occurs in
the case of $A > 1$ cosmic ray nuclei: $X_s(A,Z))\rightarrow 
Y_u(A,Z\pm1)\rightarrow X_s(A,Z)\rightarrow Y_u(A,Z\pm1)\rightarrow 
X_s(A,Z)\cdots$, where the s and u subscripts denote stability or 
instability to beta decay in the rest frame of the parent nucleus.  
After each decay the daughter nucleus has less
energy in the lab frame than the parent.

Calculating the loss in lab energy of the nucleon in a conventional beta 
decay such as $n\rightarrow p+e^{-}+ \bar{\nu_e}$ is straightforward. In 
the CM frame the proton has very little energy following the decay, and 
hence in the lab frame the nucleon loses a constant fraction $f\approx (1
- m_p/m_n)$ of its energy as a result of the decay.  For the 
energetically forbidden decay, such as $p\rightarrow n+e^{+}+ \nu_e$ the 
situation is more complex. Here for proton lab energies much above 
threshold the neutrino needs to have highly negative energies in CM
in order that its energy in the lab frame be positive, and hence the 
daughter nucleus energy can no longer be ignored in the CM frame.  The
calculation can be done as a sequence of two two-body decays: 
e.g., $p\rightarrow m(n,e^{+})+ \nu_e$ followed by $ m(n,e^{+}) \rightarrow n
+ e^+$, where in the first decay we choose only those
events having neutrinos with positive lab energy.

Figure 1 shows the $log_{10}$ of the average fractional energy loss in 
the lab frame for the nucleon in the process $p\rightarrow n +e^{+} + 
\nu_e.$  As can be seen in the figure, the fractional energy loss in this
decay is very small just above threshold, and it approaches an 
asymptotic value close to 2/3 at very high energy, where the differences 
in mass between the three final state particles becomes unimportant.

Now, at virtually all energies above threshold, the nucleon spends most 
of its time en route from the source as a neutron, because the mean free 
path for neutrons before they decay is much greater than that for 
protons except quite close to the threshold energy, where the proton mfp 
before decay becomes infinite.  Thus, the total number of steps in the 
decay chain of protons starting from the source with a given
initial energy and source distance is basically determined only by the 
{\em neutron} mfp before decay.  (This is fortunate because it means 
that the results of the calculation are quite insensitive to any 
approximations made in doing the phase space integral for the forbidden 
proton decay.)

\section*{Discussion of Results}

The results of the Monte Carlo calculation are shown in figures 2, 3 and
4 along with the data.  Following the usual practice, we show in figure
2 the log$_{10}$ of the all particle flux multiplied by a power law
($E^3$), in order to emphasize departures from this power law.  A 
reasonably good fit to the all particle spectrum can be obtained for 
$|m_{\nu}|= 0.5 eV/c^2$ (solid curve), assuming that 13 \% of
sources are ``near," with elemental abundances: 70\% A=1, 10\% A=4, 10\% 
A=14 (distributed between 5 to 19), 5\% A = 24 (distributed between 20 
to 40), and 5\% A = 56 (distributed between 41 to 90).  The solid curve 
also convolutes the Monte Carlo results with an assumed energy 
resolution $\Delta log E = \pm 0.4$ (FWHM).  As can be seen from the two 
dashed curves in figure 3, the goodness of fit worsens if the assumed
energy resolution in the convolution is $\Delta log E = \pm 0.2$ (long 
dashes) or zero (short dashes).  In these two latter cases, the fits use an
assumed tachyon mass of $|m_{\nu}|= 0.25 eV/c^2,$ and elemental
abundances: 65\% A=1, 10\% A=4, 5\% A=14, 5\% A = 24, and 15\% A = 56.

No decent fits can be obtained if $|m|$ is much greater than $0.5 eV/c^2$.
All three fits would also dramatically worsen if there were no near 
sources -- since the curves would then fall off precipitously at $E\approx
10^{19} eV.$ Thus, the flux beyond this energy appears in the model to 
come primarily from the 13 \% of sources that are``near."  The fits,
however, are insensitive to the maximum path lengths of the far source 
component.  If this is chosen as 10 Bly instead of 100 Mly, the fits are 
only slightly different from the curves in figure 2.

A convenient way to represent changes in the composition of the cosmic 
rays with energy is to plot the average $log_e$ of the atomic mass -- 
see figure 3.  The model results showing $<ln A>$ versus energy for
two different energy resolutions are only in very rough agreement with the
data in its essential features:  a rise of $<ln A>$ from the knee of the 
spectrum to a maximum in the vicinity of $10^{17}$ to $10^{18} eV$ and a 
subsequent decline to a near zero value, i.e., almost pure protons, at
$10^{19} eV$. Given the difficulty in making an experimental 
determination of the composition of cosmic rays above the knee, such 
rough agreement is probably all one could hope for.

Here is a simple explanation as to why the calculation predicts the
specific features noted in the introduction:

\begin{itemize}

\item {\bf The change in power law at $\approx 4\times 10^{15} eV$}. This is the
predicted threshold energy for cosmic ray protons to beta decay (given 
the assumed tachyon mass), and the proton component of the spectrum 
drops precipitously at this energy (to return at higher energies) -- see 
figure 2.  As the thresholds for heavier nuclei to beta decay are
reached, they are also depleted from the overall spectrum.

\item {\bf The $E^{-3}$ power law between $E= 10^{16}$ and $10^{18} eV$}
observed in the data (near horizontal slope in figure 2) is reproduced by
the model only with a proper choice of composition by atomic mass.  The 
curves shown in figures 2 and 3 used elemental abundances noted previously.

\item {\bf The minimum at $E\approx 10^{19} eV$} occurs because at this
energy the threshold for the heaviest elements to beta decay is
reached, and the spectrum becomes depleted.  However the flux 
(multiplied by $E^3$) is restored after this minimum, because at higher
energies an increasing fraction of A=1 particles from the near
source component can reach us, given their lengthened lifetime in the 
lab frame.  As noted previously, the rise in the curves after $10^{19} 
eV$ depends on the 13 \% of sources that are ``near."  Also, note that
the position of the ``ankle" (at $E\approx 10^{19} eV$) is shifted 
towards higher (lower) energies for lower (higher) assumed values of the 
neutrino mass.

\item {\bf Heavy to light transition} Just before the dip at $E\approx 
10^{19}$ the composition is very heavy -- see figure 3 -- because only
the heaviest elements are left in the spectrum at this point, since 
their thresholds have not yet been reached.  However, at the highest
energies the cosmic rays are found to be very light, because by around 
$E\approx 10^{19}$ the thresholds for all $A > 1$ nuclei have been 
reached -- and they have all been depleted from the spectrum -- while 
this energy is far enough above the threshold for A = 1 so that this
component is coming back, owing to the greatly lengthened mfp's due to 
the large dilation factor $\gamma$.

\item {\bf The abrupt change in the variation of anisotropy
amplitude}\cite{Hilas} (measured by the first harmonic of the arrival
direction of air showers) observed to occur at the knee
happens because above the threshold for proton beta decay, the A=1
component of cosmic rays is suddenly severely depleted, and hence the 
spectrum-average rigidity of the cosmic rays is reduced.

\item {\bf Cosmic rays with $E > 4\times 10^{19}$} are not blocked by
interaction with the 2.7$^0$ K cosmic background radiation (CBR), 
because as already noted, most of the flux beyond $10^{19} eV$ appears 
in our model to come from the 13 \% of sources that are ``near," so the
GZK cutoff would not apply.\cite{Takeda}  These ultrahigh energy 
particles primarily have A = 1, and as noted they spend nearly all of
their journey as neutrons.  But their neutrality is {\it not} a significant
factor in preventing their interaction with the CBR, since neutrons 
would be expected to photoproduce pions with essentially the same cross section as
protons.

\end {itemize}

Thus, in summary, our calculation of the cosmic ray spectrum above the 
knee follows naturally from the hypothesis that the electron neutrino is 
a tachyon with mass $m_{\nu} \approx 0.5 eV/c^2,$ and it is consistent with
some other observations of the cosmic rays.  However, any model has its
problems and challenges, and we now turn to some of those.

\section*{Potential problems with the model}

\begin{itemize}

\item {\bf conventional explanations exist} for some of the regularities 
we have noted, and plausible mechanisms exist to account for the 
production of the component of the spectrum believed to be galactic in 
origin.  However, few conventional explanations predict numerical values 
for the position of the knee and ankle, and many of the models have both 
ad hoc elements and many free parameters.  Moreover, some of the 
features we have noted have no conventional explanation, and some 
represent a very severe test of all conventional models -- particularly 
the {\it abruptness} of the change in slope at the knee and ankle of the 
spectrum.\cite{Erlykin}

\item {\bf statistical significance}  Some of the particular 
regularities our model predicts are not terribly significant 
individually.  Moreover, we must consider the element of ``informed 
choices," i.e., choosing values of parameters that yield the best fit to 
the data, rather than making predictions before having seen the data.  
For this reason it is not possible to assign any probability of the 
model fitting the data simply on the basis of chance.


\item {\bf composition independent of energy}  As noted previously, our 
model assumes an elemental composition at the source that is independent 
of energy, and then deduces the composition as seen on Earth, based on 
what gets depleted, or shoved down to lower energies through a chain of 
decays.  It is highly unrealistic to suppose that the composition at the source
is energy-independent, but by making this assumption we are merely 
limiting the number of free parameters in the model, making it more 
difficult to achieve a fit, not less.


\item {\bf Other models can account for the absence of a GZK cutoff} 
Various suggestions have been made to explain why cosmic rays with 
energies above the conjectured GZK cutoff ($E\approx 4\times 10^{19} 
eV$) apparently fail to be significantly degraded in energy by 
interaction with the CBR.  One category of suggestions is that these 
cosmic rays are a new type of particle which interacts with CBR photons 
less strongly than nucleons or nuclei.\cite{Farrar}  The hypothesis of 
remote sources (though not necessarily that of new types of particles) 
would receive additional support if it is found that they do point back 
to known distant radio quasars, as Farrar has suggested.\cite{Farrar}  
Another suggestion put forth by Coleman and Glashow is that Lorentz 
symmetry is slightly broken.\cite{Coleman}  This would have the effect 
of kinematically forbidding interactions between cosmic ray protons and 
CBR photons, when the former have an energy above a certain value.  
Despite these alternative hypotheses for explaining the avoidance of a 
GZK cut-off, it would seem that the least exotic hypothesis is that the 
sources of cosmic rays with $E > 4\times 10^{19} eV$ simply are closer 
than a few dozen Mpc (as our model requires).


\item {\bf no known sources with a single power law.}  No conventional 
mechanisms are known that have a $E^{-2.67}$ power law spanning over ten 
decades, and it is not even clear what unconventional sources might be a 
candidate.  Of course, there are no known sources in the conventional 
theory of cosmic rays at the highest energies either. Topological 
defects -- including magnetic monopoles, cosmic strings, domain walls, 
etc. -- have been suggested as sources for the highest energy cosmic 
rays.\cite{Bhatt}  But they have not been proposed to account for the
lower energy region, which are believed to originate from shocks driven 
into the interstellar medium by supernova explosions in the generally 
accepted conventional view.  However, even for the lower energy region, 
the evidence supporting the conventional view of supernova shock waves 
as the source is only circumstantial, i.e., SN's have enough energy for 
steady state cosmic ray production, and they would have a power law 
index at the source that is flatter than what is observed in the 
arriving flux, as expected.  But, direct confirmation of the 
conventional theory of supernova shock sources does not yet exist, e.g., 
high energy gamma rays from pions produced during the acceleration 
process.  One exotic possibility for sources has been proposed by Kuzmin 
and Tkachev:\cite{Kuzmin} the decay of supermassive long-lived particles produced in
the early universe from vacuum oscillations during 
inflation.  One advantage of this possibility from the
point of view of our model is that such sources could be a considerable 
fraction of cold dark matter, and hence could be prominent in the Milky 
Way galactic halo, and therefore relatively nearby.  Yet, they would 
also be relatively isotropic, as seems to be the case for the limited 
number of events so far seen at the highest energies.


\item {\bf source spectrum.} Choosing the source spectrum to match the 
observed $E^{-2.67}$ power law below the knee probably is another 
unrealistic feature of the model.  A more realistic model would have 
included a source spectrum less steep than $E^{-2.67}$ in order to include
other energy loss processes besides those included here.  Without such 
processes it would be difficult to account for the observed primary to 
secondary ratios.  However, the model is not intended
to be an all-encompassing explanation of every aspect of the cosmic ray 
spectrum, but rather an example of how a tachyonic electron neutrino can 
account for many of its features.

\item {\bf no observed neutron component seen in the cosmic ray spectrum.}
Although no neutron component has yet been observed in the cosmic ray flux,
current techniques based on air shower measurements would not 
distinguish between proton-induced cascades and those initiated by
neutrons.  In fact, based on anisotropy data, Tkaczyk\cite{Tkaczyk} has 
estimated that the neutron component could be as high as 20 \% in the
$10^{16} - 10^{18}$ eV energy region if the neutrons come from sources 
in the galactic disk.  (In our model, the chain of decays $p\rightarrow n\rightarrow
p\rightarrow n\rightarrow p\cdots$ only occurs at energies above the 
threshold for proton decay, so neutrons would not be seen below this 
energy, given their mean free path before decay.)

\end{itemize}

\section*{Possible Confirming Tests of the Hypothesis}

As already noted, most tritium beta decay experiments have consistently 
reported negative values for ${{m_\nu}_e}^2.$  The seven tritium beta 
decay experiments used by the Particle Data Group in 1998\cite{PDG} to 
find a value for ${{m_\nu}_e}^2$ all report negative values.  Two of 
these experiments report values that are negative by over four standard 
deviations, but they are inconsistent with each 
other:\cite{Belesev,Stoeffl} ${{m_\nu}_e}^2 = -22\pm 4.8$ and $-130\pm 
20\pm15 eV^2/c^4$ Regretably, the value we have used here 
$|{{m_\nu}_e}|\approx 0.5 eV/c^2$ is too small to be consistent with the 
negative values reported for ${{m_\nu}_e}^2$ in either of these 
experiments.  Moreover, the tritium results have been explained in terms 
of either experimental anomalies,\cite{Barth,Lobashev,Rizek} final state 
interactions, or new physics\cite{Stephenson} -- though a few authors 
have attributed them to tachyonic 
neutrinos.\cite{Rembielinski,Ciborowski}

Assuming that the electron neutrino really were a tachyon, could future 
tritium beta decay experiments test for values of ${{m_\nu}_e}^2$ as 
small as $0.25 eV^c/c^4$?  The current systematic and statistical errors 
on $m^2$ are over an order of magnitude smaller than that value, so 
without new types of instruments the answer is probably not.\cite{Bonn}  
But, apart from the issue of the limits of experimental precision, and the
need to better understand the basis of the negative ${{m_\nu}_e}^2$ 
values found in past experiments, it is also important that 
experimenters avoid the position that ``the negative value for the best 
fit of ${m_{\nu}}^2$ has no physical meaning."\cite{Belesev,Stoeffl}  Of 
course, experimentalists are not alone in believing negative mass$^2$ 
particles are unphysical.  For example, Hughes and Stephenson have 
argued that if the neutrino really were a tachyon, there would be no 
endpoint to the electron energy spectrum in tritium beta decay, and a 
large amount of phase space would exist for neutrinos of arbitrarily 
large negative energies.\cite{Hughes}  But, their result ignores the 
reinterpretation principle, according to which negative energy emitted 
tachyons are only part of the kinematically allowed region for tritium 
decay if they have positive energy in the lab frame.

If one is open to the idea that neutrinos could, in fact, be tachyons it 
is natural to ask why one should put any more faith in the mass value 
obtained from the fit to the cosmic ray spectrum ($|{{m_\nu}_e}|\approx 
0.5 eV/c^2$) than the much larger values found in tritium beta decay 
experiments?  One answer is that the only statistically significant 
negative values found in tritium beta decay experiments are inconsistent 
with each other, and have been attributed to a number of plausible 
causes having nothing to do with tachyons.  A second answer is that if 
any of the values from tritium beta decay experiments represent masses 
of real tachyons, then the knee of the cosmic ray spectrum would have to 
occur at an energy one or two decades lower than is observed, because 
the threshold energy for proton beta decay varies inversely with 
$|{{m_\nu}_e}|.$ Alternatively, if the tachyon mass found from the 
cosmic ray spectrum fit is correct that only means that the values
reported in the tritium experiments arise from causes other than real 
tachyons.

Are there other places one might look for confirmation of the hypothesis 
that the electron neutrino is a tachyon?  Neutrino oscillation 
experiments, being sensitive to differences in $m^2$ cannot reveal 
whether any neutrino flavors have negative $m^2.$  Caban, Rembielinski 
and Smolinski have suggested that the tachyonic neutrino hypothesis may 
explain the problem of the missing solar neutrinos.\cite{Caban}  But, of 
course, the missing solar neutrino problem has already been explained in 
terms of neutrino oscillations and other less radical ways than the 
hypothesis of tachyonic neutrinos.  The possibility of testing whether 
${{m_\nu}_e}^2 < 0$ based on observing neutrinos from some future 
supernova remains a possibility, but it is uncertain if a value of 
$|{m_\nu}_e|$ as small as $0.5 eV/c^2$ could be clearly distinguished 
from zero when the current upper limit\cite{PDG} based on SN1987A is 
$|{m_\nu}_e| < 15 eV/c^2.$

There is, however, one unambiguous test of the tachyonic neutrino 
hypothesis involving a cosmic ray neutron flux.  Graphs of cosmic ray 
flux times E$^3$ (see figure 2) may be good for spotting changes in the
flux power law, but they can be extremely misleading in other respects.  
For example, there really is no ``ultraviolet catastrophe," as figure 2
seems to show, after one removes the $E^3$ multiplier.  In addition, it 
might seem from the dotted curve in figure 2 that the A = 1 contribution
to the flux drops precipitously right after the threshold for proton 
decay, and does not return until around $E\approx 10^{19} eV.$  In fact, 
however, the small bump in figure 2 just before the precipitous drop is
actually a rather large bump when one plots the flux without multiplying 
it by $E^3,$ -- see figure 4, which shows both the neutron flux, and its
fraction of the total plotted on a linear rather than a log scale.

Thus, a most distintive prediction of the model is a spike of
neutrons just above the threshold energy for proton beta decay.  The 
position and height of the neutron spike does of course depend on the 
value assumed for the neutrino mass -- $|m| = 0.25 eV/c^2$ in figure 4.
Doubling that value would halve the energy at which the spike occurs, 
and increase its height eightfold.  The accumulation of a spike of 
neutrons just above the threshold energy is a consequence of the 
fractional energy loss of the nucleon becoming very small as the 
threshold is approached from above -- see figure 1.

Experimentally, it may be impossible to distinguish individual cosmic 
ray neutrons from protons in the region of the knee of the spectrum, 
based on air shower measurements.  But, there is one unambiguous 
difference: unlike protons or nuclei, neutrons point back to their 
sources in this energy region.  Hence, multiple events coming from the 
same directions should be a clear indicator of neutrons.  Moreover, 
given the neutron lifetime, the mean free path before decay at an energy 
of $10^{16} eV$ is only about 6000 ly -- much too close for many sources 
in any conventional model.  As figure 3 shows, neutrons should also be 
seen as a large component of the (much smaller) flux at energies above 
$10^{19} eV.$  However, if neutrons were seen at these energies, their 
presence could well be the result of sources closer than 6 Mly, and they 
would, therefore, have little value in confirming the hypothesis of 
tachyonic neutrinos.

Interestingly, there is one other hypothesis that has been suggested to 
account for the knee of the cosmic ray spectrum that also predicts that 
neutrons should be a component of the cosmic ray flux beginning just 
above the knee of the spectrum.  Wigmans has recently suggested that 
massive relic neutrinos might be gravitationally captured and 
concentrated around cosmic ray sources such as neutron 
stars.\cite{Wigmans}  If the relic neutrinos could be concentrated to a 
sufficient degree, cosmic ray protons passing through these stellar 
``neutrino atmospheres" would lose energy though the inverse beta decay 
process: $p + \bar{\nu} \rightarrow n + e^+.$  If the relic neutrino has 
very little energy and a very small mass, the energy threshold for the 
reaction is almost identical for the threshold for $p \rightarrow n + 
e^+ +\nu,$ where $\nu$ is a tachyon.\cite{Wigmans}  But, of course, if 
Wigmans' hypothesis were correct, few of the neutrons produced by cosmic 
ray protons just above the knee of the spectrum could reach us,
given a mean free path before decay of only 6000 ly.  In contrast, with 
the tachyonic neutrino hypothesis, neutrons can reach us even if the 
sources are at much greater distances, since a chain of decays occurs: 
$p\rightarrow n\rightarrow p\rightarrow n\rightarrow p\cdots$ resulting 
in an observed flux of neutrons piling up just above the threshold for 
proton decay.

This is not the place to attempt to convince the reader that the 
paradoxical properties of tachyons -- especially their apparent 
violation of causality -- does not make their existence absurd.  Let it 
only be said that the plausibility of tachyons should hinge more on 
whether they are capable of absurd possibilities, e.g., sending messages 
back in time, than on any theoretical formulation of a definition of 
causality, and that the various absurd possibilities sometimes 
attributed to tachyons can be dismissed by reinterpreting emitted 
(negative energy) tachyons as being absorbed positive energy ones, as 
various authors have demonstrated.\cite{Bilaniuk,Feinberg,Chodos85}
If tachyons do really exist, either nature would have found a way of 
eliminating absurd possibilities, or else physicists would have to 
readjust their thinking as to the boundaries of the absurd.



\section*{Acknowledgements}
The author wishes to thank his colleague, Robert Ellsworth, for making
some helpful suggestions regarding possible objections to the model, and 
for providing invaluable advice, assistance, and moral support.  He
also wishes to thank Alan Chodos and Len Ozernoy for their support, and 
their comments on a draft of this paper.

\newpage

\begin{figure}[htb]
\begin{center}
\leavevmode
\epsfysize=3.5in
\epsffile{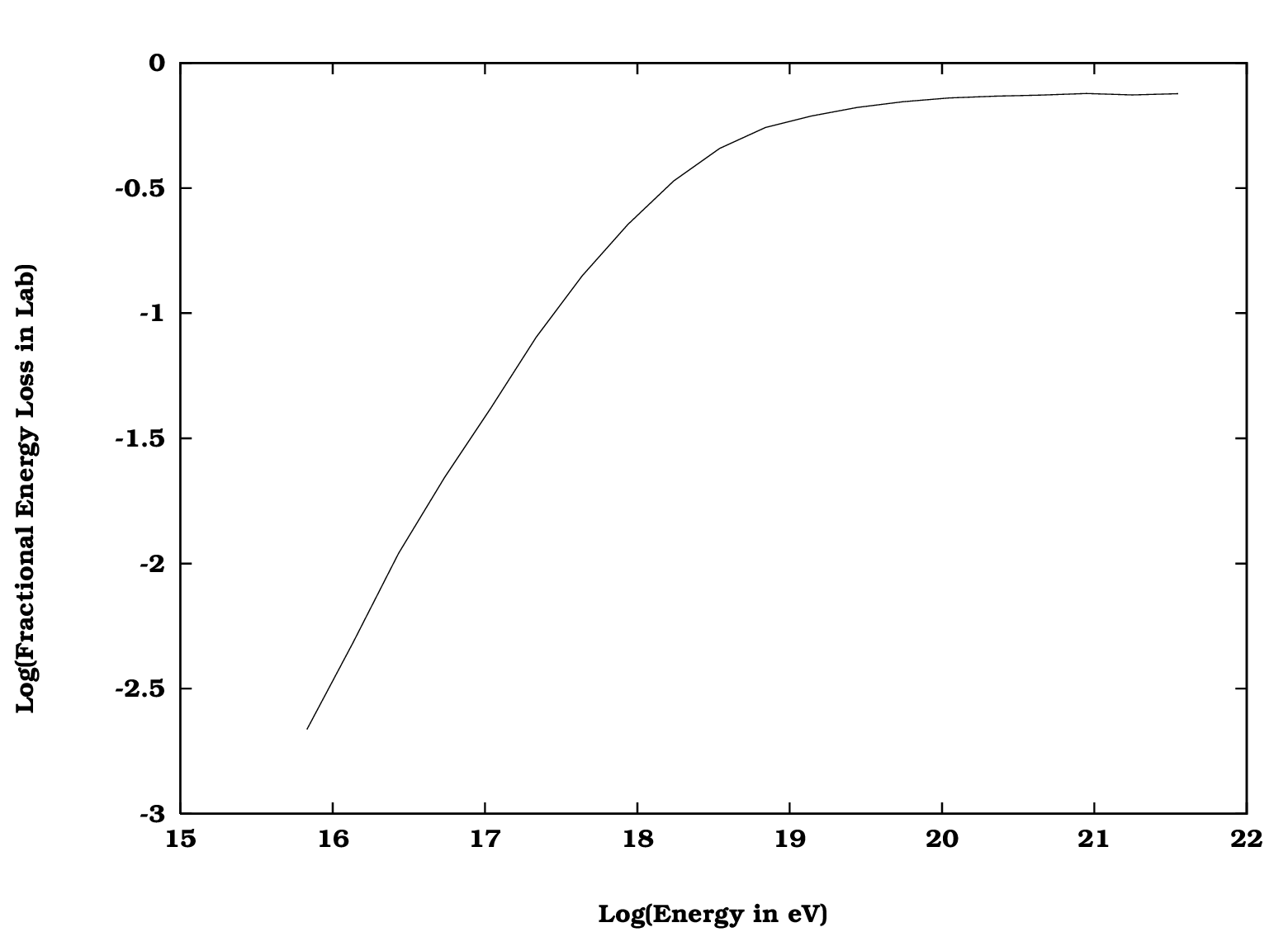}
\caption{Log$_{10}$ of the average fractional energy loss of the nucleon 
in the decay $p\rightarrow n+e^{+}+\nu_e,$ assuming a neutrino mass $|m| 
= 0.5 eV/c^2$ for proton energies above the threshold.}
\end{center}
\end{figure}

\begin{figure}[htb]
\begin{center}
\leavevmode
\epsfysize=3.5in
\epsffile{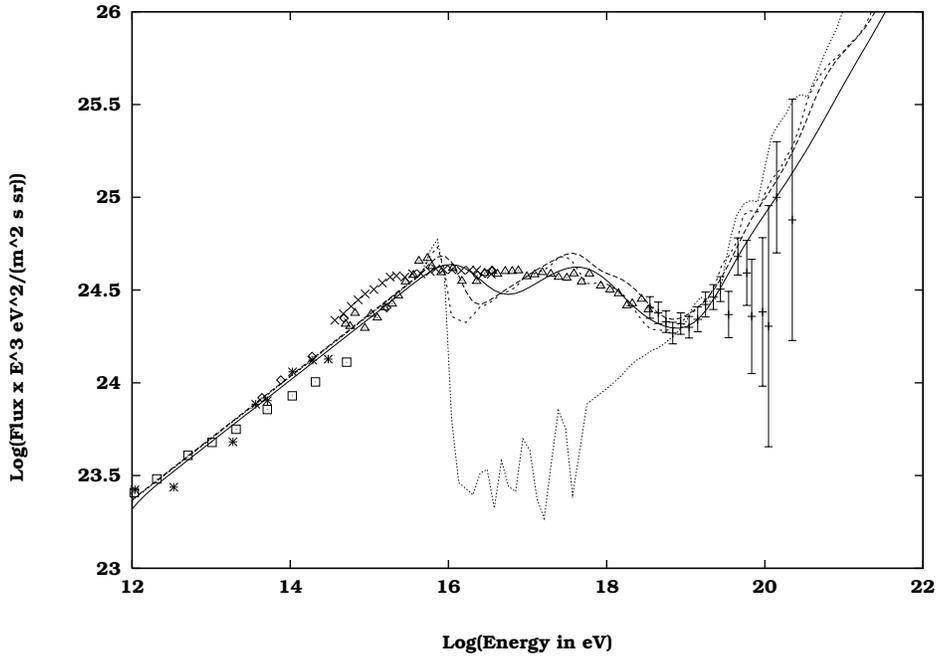}
\caption{Solid curve shows the prediction of the model for the cosmic 
ray flux $(\times E^{3})$ assuming a tachyon mass $|m| = 0.5 eV/c^2$, 
with convolution, assuming an energy resolution of $\Delta log E=\pm
0.4$.  The two dashed curves show fits with a tachyon mass of $|m| =
0.25 eV/c^2$: short dashed curves assumes no convolution to account for 
energy resolution, and the long dashed curve uses an energy resolution 
$\Delta log E=\pm 0.2$. The dotted curve shows the spectrum with A = 1 
only using $|m| = 0.25 eV/c^2$.  All curves assume 13 \% near sources
with mass compositions noted in the text.  Points are the data from: 
JAYCEE (diamonds), AGASA (with error bars), Aoyama-Hirosaki (squares), 
Tibet (crosses), Akeno 1km$^2$ array (diamonds), Proton Satellite 
(asterisks).}

\end{center}
\end{figure}

\begin{figure}[htb]
\begin{center}
\leavevmode
\epsfysize=3.5in
\epsffile{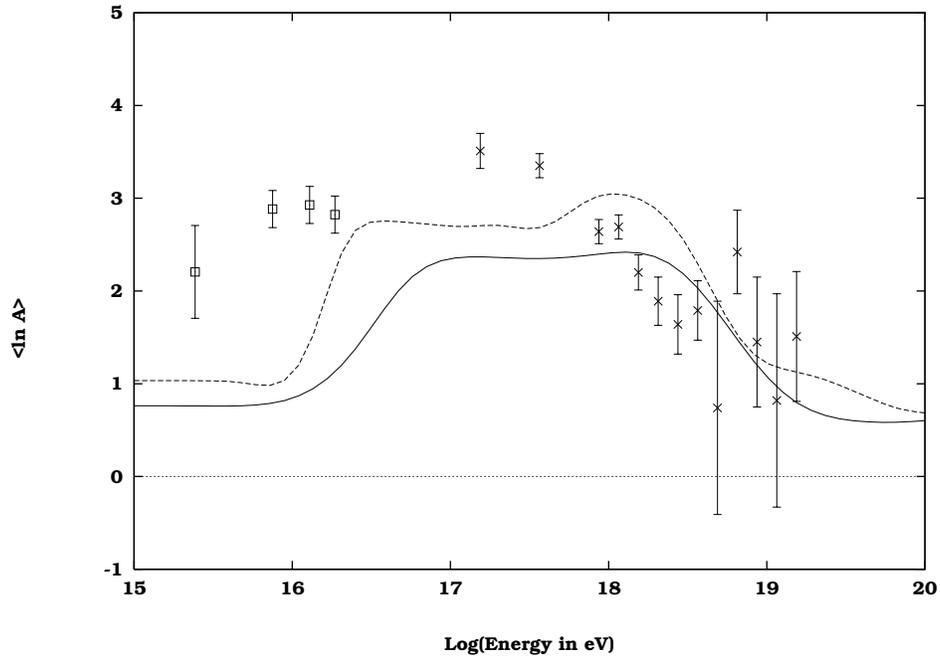}
\caption{Prediction of the model for the cosmic ray composition ($<ln 
A>$) as a function of particle energy.  Solid and dashed curves makes 
the same assumptions of values of the tachyon mass, composition, and the 
percentage of ``far" cosmic ray sources as the solid and dashed curves 
shown in figure 3.  Data points with a horizontal line segment are from 
JACEE (1995), squares are BASJE (1994), and crosses are Fly's Eye 
(1993).  For the data from the stereo Fly's Eye (squares), we have made 
a linear interpolation in $<ln A>$ using their elongation rate data to 
deduce values for $<ln A>$}

\end{center}
\end{figure}

\begin{figure}[htb]
\begin{center}
\leavevmode
\epsfysize=3.5in
\epsffile{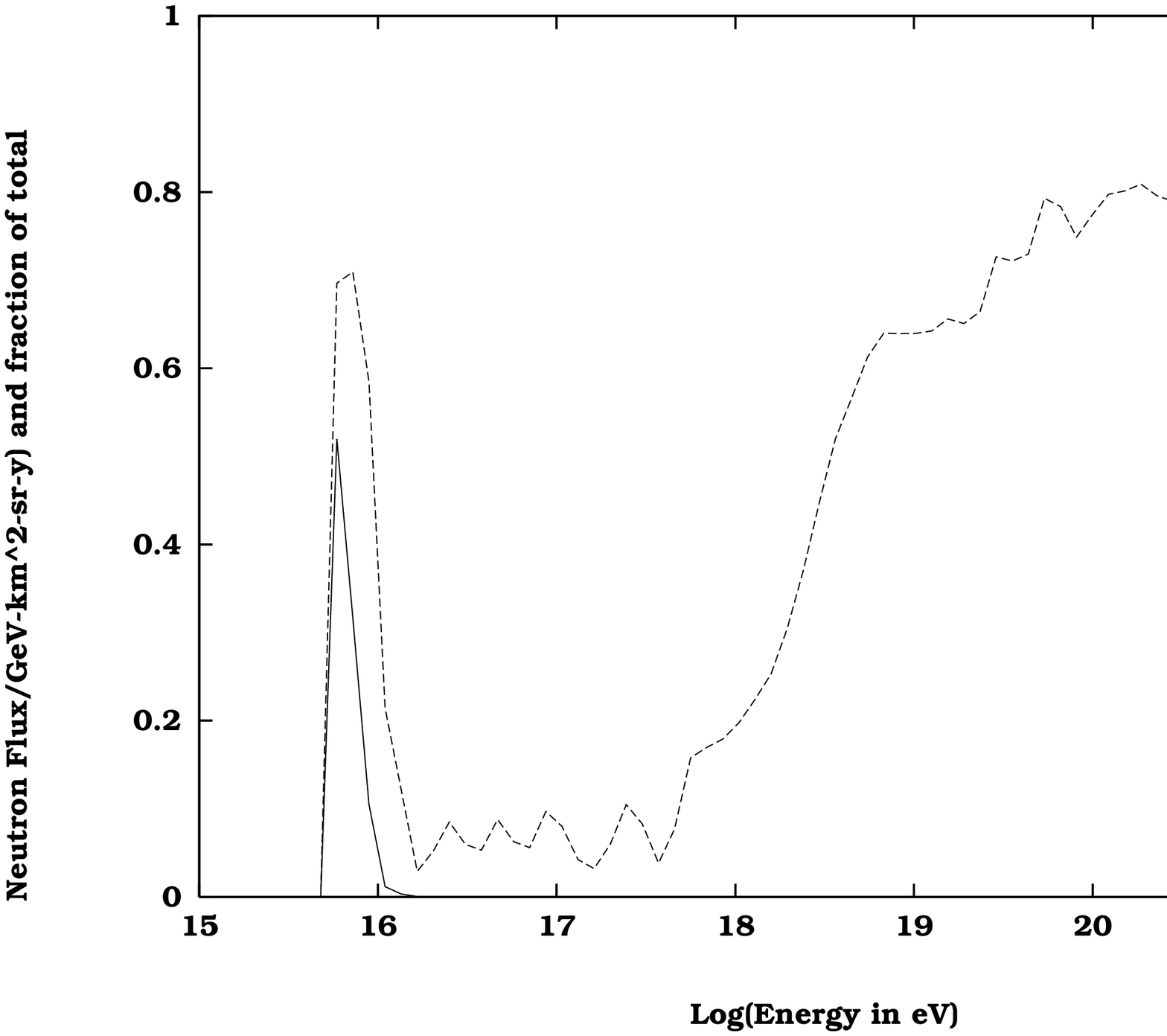}
\caption{Solid curve shows the prediction of the model for the cosmic 
ray flux of neutrons, assuming a tachyon mass $|m| = 0.25 eV/c^2$, 13\%
near sources, with no convolution to account for finite energy 
resolution.  The dotted curve shows the fraction the neutron flux is of 
the total flux.  The flux has been expressed in units that allow the same
vertical scale to be used for both curves.  Doubling the tachyon mass
would shift the neutron peak to half the energy, or shift log E downward 
by 0.3, and increase the peak height eightfold.}

\end{center}
\end{figure}



\begin{thebibliography}{99}
\bibitem{tritium} The tritium beta decay experiments reporting negative 
${m_{\nu}}^2$ values, which could be regarded as a positive signal, were not
specifically designed to look for tachyons.
\end{thebibliography}
\end{document}